\documentstyle[twocolumn,natb_209,epsf]{article}
\def\hMpc{\mbox{h$^{-1}$ Mpc}}

\def\deg{\ifmmode^\circ\else$^\circ$\fi}    
\def\SS{Sect.~}
\def\ltapprox{\,\lower.6ex\hbox{$\buildrel <\over \sim$} \, }
\def\ttimes{{\scriptstyle \times}}
\def\exunit{{\bf e_x}}  
\def\eyunit{{\bf e_y}}  
  
\renewcommand\citep[1]{(\citealt{#1})}
\newcommand\citepf[1]{(\citealt*{#1})}    

\def\mycaptionfont{\protect\footnotesize} 

\newcommand\zzz[2]{#2}  

\def\ftopofour{
\begin{figure}
\centering 
\centerline{\epsfxsize=8cm
\zzz{\epsfbox[0 0 522 765]{"`gunzip -c topo4banyan.eps.gz"}} 
{\epsfbox[0 0 522 765]{"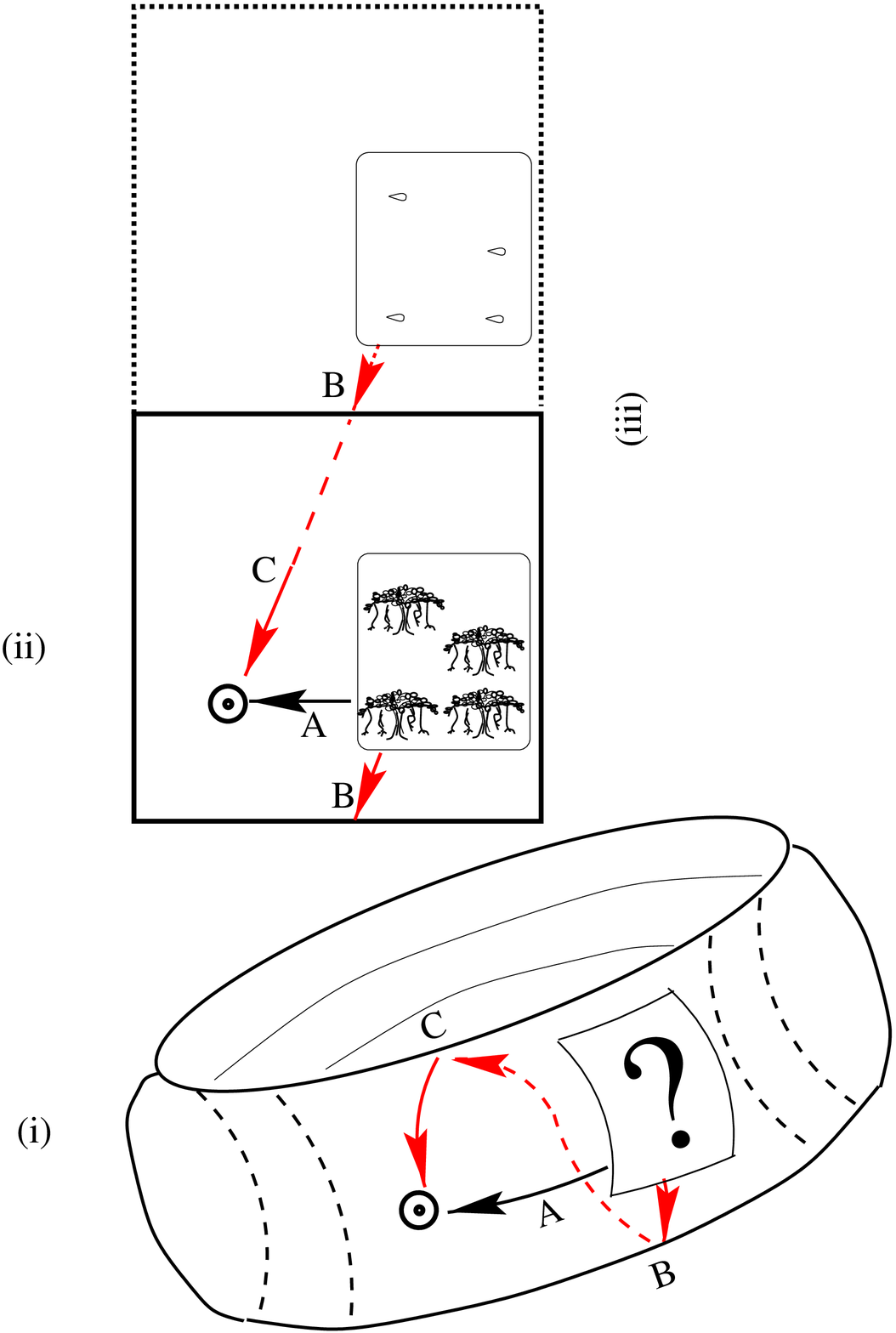"}}  }
\caption[]{ \mycaptionfont
Schematic illustration to show different ways of thinking of
a 2--torus, $T^2$, and how photons may come to the observer 
from astrophysical objects by multiple paths. 
The lower part
of the figure shows (i): $T^2$ `stretched' into $R^3$, Euclidean
3--space (and then projected into the plane of the paper), and 
possible paths from the object `?' to the observer either via
the short path A or via the longer path BC. The lower rectangle of the
upper part of the 
figure shows (ii): $T^2$ as a rectangle with identified sides, 
where again two different paths to the observer, A and BC, are shown
from the astrophysical object, now shown as a `configuration' of 
banyan trees (to use a familiar object).
The two rectangles together show part of (iii): since the path BC
is longer than A, the observer in fact will see the astrophysical
objects as they were a long time in the past, e.g. banyan trees 
would be seen as seeds (if they lived several Gigayears and were
visible at cosmological distances!), and could be assumed to be
present in an apparent position in another copy of the original 
rectangle, without introducing any error into the calculation of
geodesics, time lags, etc. Version (iii) also provides the simplest
way to correctly calculate the angles (perspectives) 
of viewing an object, which change (in general) between multiple
images. For example, if a left-front-right-back 
orientation were added to the figure, we could say that the `banyan trees'
are viewed more or less from their `left-hand' sides here, but that
their younger versions, i.e. seeds, are viewed more or less 
`from the front'.
}
\label{f-topo4}
\end{figure}
} 

\def\fhowcirtwo{ 
\begin{figure}
\centering 
{\epsfxsize=8cm
\zzz{\epsfbox[100 30 600 491]{"`gunzip -c howcir2.eps.gz"}}
{\epsfbox[100 0 600 491]{"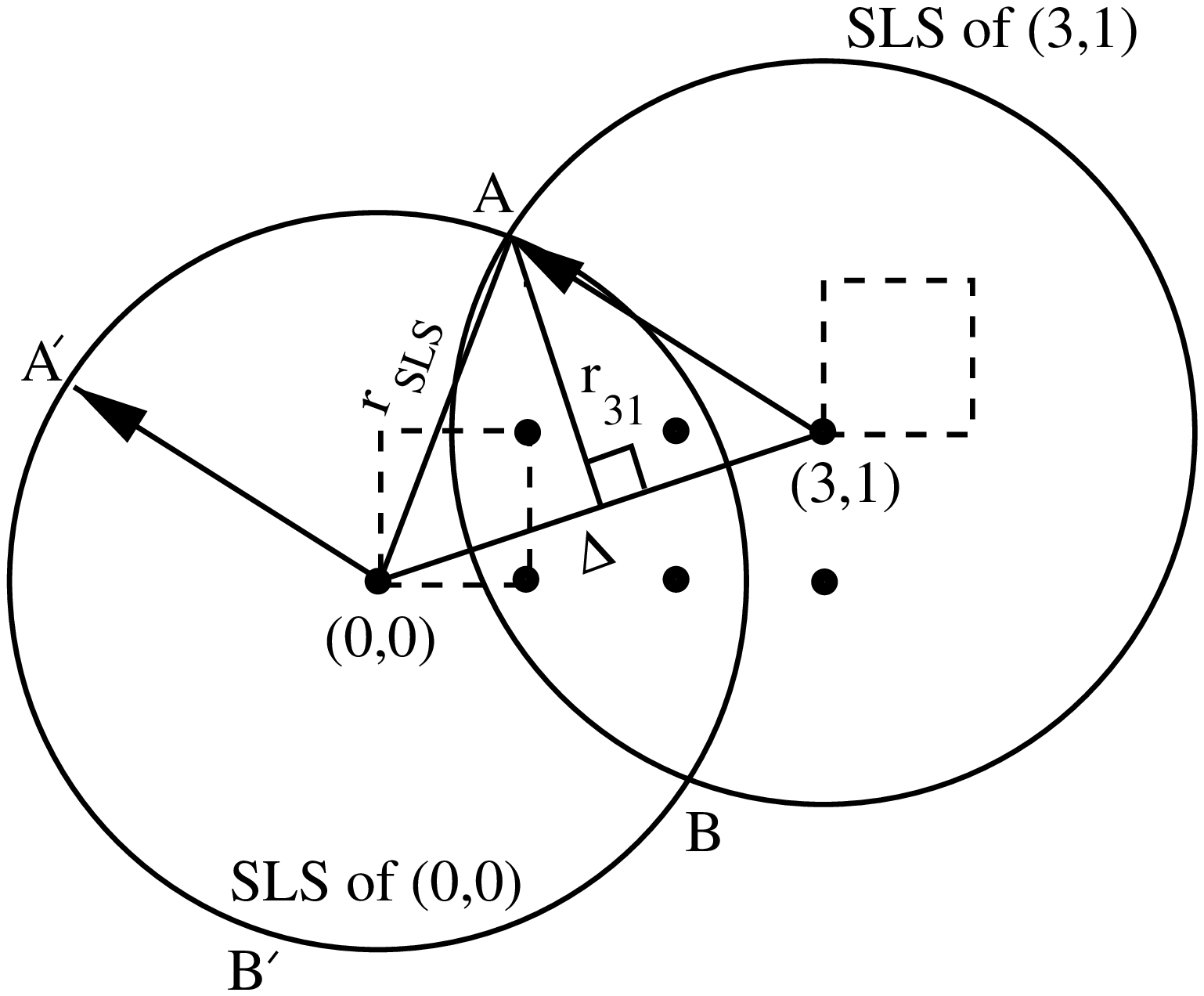"}}  }
\caption[]{ \mycaptionfont
Illustration of `identified circles principle', shown with one
dimension less, i.e. with two 1-spheres (circles) that intersect in a
0-sphere (pair of points), instead of two 2-spheres that intersect in
a 1-sphere (circle). The two 1-spheres, or surfaces of last scattering
(SLS), of radius $r_{\mbox{\small SLS}}$,
are centred at the observer at $(0,0)$ and at a copy of the
observer at $(i=3,j=1)$ in the covering space, where $i$ and $j$
represent numbers of copies of the fundamental polyhedron (rectangle
in 2-D, rectangular prism in 3-D, shown as a dashed curve).  Their
intersection consists of the two points $A$ and $B$ (in 3-D this
intersection is a full circle).  The two observers looking at $A$ and
$B$ from `opposite' sides are equivalent to one observer, at $(0,0)$,
looking at $A$ and $B$ in one direction, and at $A'$ and $B'$ in
some other direction, defined by the tiling pattern of the fundamental
polyhedron and coordinate location of the observer's copy. In 3-D,
the two identified circles for a single observer defined by copy
$(i,j)$ are the extensions of the 0-sphere $\{A,B\}$ and 
the 0-sphere $\{A',B'\}$ into 1-spheres (circles). The radii of
the identified circles are $r_{31}$, and the two circles are
separated by the observer-observer copy distance, labelled $\Delta$
in the figure. Black dots are used to show some other possible 
positions of a copy of the observer, each of which defines a distinct
pair of identified circles.
}
\label{f-howcir2}
\end{figure} 

} 

\def\fcircles{ 
\begin{figure}
\centering 
{\epsfxsize=8cm
\zzz{\epsfbox[46 26 479 390]{"`gunzip -c circ_gd.ps.gz"}}
{\epsfbox[46 26 479 390]{"circ_gd.ps"}}  }
{\epsfxsize=8cm
\zzz{\epsfbox[46 26 479 390]{"`gunzip -c circ_bd.ps.gz"}}
{\epsfbox[46 26 479 390]{"circ_bd.ps"}}  }
\caption[]{ \mycaptionfont
Two examples of would-be identified circles shown 
as temperature fluctuations $\delta T/T$
in the cosmic microwave background as observed by 
the COBE satellite. The upper panel shows $\delta T/T$ around 
a pair of circles 
implied by a model found to be consistent with the four-year COBE data,
the lower panel shows $\delta T/T$ around a pair of circles for 
a model found inconsistent with the full data set. 
The circles do not cover $360\deg$; this is due to exclusion of 
galactic latitudes with $|b^{\mbox{{\sc II}}}| < 20\deg$.
Thick lines and thin lines indicate $\delta T/T$ and 
$\delta T/T \pm \Delta (\delta T/T)$ (uncertainties) respectively,
assuming that 30\% of the total signal is due to
the integrated Sachs-Wolfe effect, added to the noise in quadrature.
For a correct model, the $\delta T/T$ values around matched circles 
should be equal within the statistical limits of the uncertainties, 
assuming that the values are primarily due to the na\"{\i}ve
Sachs-Wolfe effect. The horizontal axes are labelled in lengths 
$r_{ij} \theta$ around the circles of 
radii $r_{ij}$, for $(\Omega_0=0.3, \lambda_0=1-\Omega_0)$.
The two models are `2-tori', i.e. $T^2 \ttimes R$, of size one-tenth
of a horizon diameter, but oriented differently in the two cases.
The labels $(i,j)$ indicate which observer-observer copy pair defines
the circle pair (see Fig.~\protect\ref{f-howcir2}).
Each circle lies in a plane halfway between the observer 
and her topological image at 
$i (2R_H/10) \exunit + j (2R_H/10) \eyunit$ 
and its matching circle is at $(-i,-j),$ 
where $\exunit$ and $\eyunit$ are unit vectors in the two short directions
of $T^2 \ttimes R$.
The distance between a circle and its match is
indicated here by $\Delta.$ 
}
\label{f-circles2}
\end{figure} 
} 

\begin{document}


\title{The Topology of the Universe}


\author{Boudewijn F. Roukema, Inter--University Centre for 
Astronomy and Astrophysics\\Post Bag 4, Ganeshkhind, Pune, India\\
{\em boud@iucaa.ernet.in} }

\date{To appear in {\em 
Bulletin of the Astronomical Society of India} (Sep 2000)}

\maketitle

\begin{abstract}
The Hilbert-Einstein equations are insufficient to describe the
geometry of the Universe, as they only constrain a local geometrical
property: curvature. A global knowledge of the geometry of space, if
possible, would require measurement of the topology of the
Universe. Since the subject was discussed in 1900 by Schwarzschild,
observational attempts to measure global topology have been rare for
most of this century, but have accelerated in the 1990's due to 
the rapidly increasing amount of observations of non-negligible
fractions of the observational sphere. A brief review of basic
concepts of cosmic topology and of the rapidly growing gamut
of diverse and complementary observational strategies for 
measuring the topology of the Universe is provided here. 
\end{abstract}

\section{Introduction}

Is it possible to observe the whole Universe? Is the object
studied in what is claimed to be observational cosmology really 
all of space or just a tiny bit of space?

In order to answer these questions, 
it is necessary to measure the mathematical properties of
{\em local geometry} (such as the curvature) and {\em global geometry}
(such as the topology), which together 
describe the `shape'  and size of space, under the assumption
that the curvature is nearly constant everywhere in space. 
 Since 
a century ago, Schwarzschild (1900), de Sitter (1917), Friedmann
(1924) and Lema\^{\i}tre (1958) have realised that the spatial 
part of our Universe could correspond to a space (a 3--manifold) 
which may have either a non--zero
curvature and/or a non--trivial topology.

The measurement of these properties (one local and the other global)
from surveys obtained at
telescopes of different sorts, such as the GMRT, 
the AAT, the VLT, XMM, MAP and Planck Surveyor, should enable us
to find out if our cosmological observations are global in the
sense of measuring the whole of space, or whether they simply 
measure a tiny fraction of the Universe: our observable sphere.

Tests for measuring curvature or topology are dependent 
to differing extents on assumptions of the 
cosmological model adopted.
Most tests are evaluated in terms of the most popular model,
i.e. the `Hot Big Bang'
model, or in other words, the perturbed Friedmann--Lema\^{\i}tre model,
but as long as the cosmological expansion interpretation of redshifts
is retained, many of the tests involving `standard candles' or 
`standard rulers' should also be valid for the quasi steady state
cosmology model \citepf{HBN93}.

In order to aid the non--specialist, some reminders on curvature
and topology are provided in \SS\ref{s-definegeom}. 
The application of these geometrical concepts to the standard hot Big Bang
model, to extrapolations of the standard model and to the
quasi steady state cosmology model are presented in \SS\ref{s-models}.

What do the observations tell us? 
Serious observational work with what 
may be hoped to be sufficiently deep surveys to 
determine the global geometry of the Universe have only just 
started in the last decade, and the race is on to obtain the first
significant results. A brief glance at the various strategies using
different astrophysical objects or radiation sources and tentative
results is described in \SS\ref{s-obsvns}.

Comoving coordinates are used to describe space throughout this review.

\label{s-definegeom}
\section{Some basic geometry: curvature and topology}
In the standard Friedmann--Lema\^{\i}tre cosmology, the model
of space--time is locally based on the Hilbert--Einstein equations, 
where local geometry (curvature) 
is equated to local physical content (density) of the Universe.
Such a space--time has spatial sections (i.e. hypersurfaces at
constant cosmological time) which are of constant curvature.

In order to intuitively understand curvature, it is useful to 
use a two--dimensional analogy. An example of a flat, or curvature zero,
two--dimensional space is the Euclidean plane ($R^2$). 

\ftopofour

Two examples of non--zero (but constant) curvature two--dimensional 
spaces (or surfaces) are the sphere ($S^2$) and the hyperboloid
($H^2$), which are of positive and negative curvature respectively.

These three spaces are {\em simply connected}, i.e. any closed 
loop on their surfaces can be continuously contracted to a point.
This would not be the case if there was a `handle' added to one of
these surfaces, because in that case any loop circling the hole
of the handle (for example) would not be contractible to a point.
A space for which there exist non--contractible loops is called 
{\em multiply connected}.

An example of a flat, multiply connected space is the {\em flat torus} 
($T^2$). There are three different ways to think of this space, 
each useful in different ways, explained further below and
in Fig.~\ref{f-topo4}:
\begin{list}{(\roman{enumi})}{\usecounter{enumi}}
\item as a sort of `doughnut' shape by 
inserting it in a three--dimensional Euclidean space, but retaining
its flat metric (rule for deducing distances between two close points),
\item as a rectangle of which one physically identifies opposite
sides, or
\item as an apparent space, i.e. as a tiling of the full Euclidean
plane by multiple apparent copies of the single physical space.
\end{list}

The polygon (or in three dimensions, the polyhedron) of (ii) 
is termed the {\em fundamental polyhedron} (or {\em Dirichlet domain}).
The apparent space (iii) is termed the {\em universal covering space}, 
or {\em covering space} for short. Representation (i) is not 
generally useful for analysis of observations.

One can shift between (i) and (ii) by cutting (i) the `doughnut' shape
twice and unrolling to obtain (ii), or by rolling and sticking together
opposite sides of (ii) the rectangle in order to obtain (i).
These operations help us to see why $T^2$ and $R^2$ are 
{\em locally} identical, i.e. both have curvature zero, 
since the former can be constructed from
the latter by cutting a piece of the latter and pasting, but
that {\em globally} they are different, since $T^2$ has a finite
surface area (is `compact'), without having any edges, but 
$R^2$ is infinite.

This can now be put in a cosmological context (imagining a two--dimensional
universe), by thinking of a photon which makes several crossings of
the torus $T^2$, i.e. of a universe. 
In the three ways of thinking of $T^2$, this 
can be thought of as (i) looping the torus several times, 
(ii) crossing the rectangle, 
say, from left to right several times or (iii) in 
the covering space (apparent space), crossing many copies of 
the rectangle before arriving at the observer.

Of course, this is only possible if the time needed to cross 
the rectangle is less than the age of the universe.

In three dimensions, the three simply connected 
constant curvature spaces corresponding
to those listed above are the 3--D Euclidean space, $R^3$, the
hypersphere ($S^3$) and the 3--hyperboloid ($H^3$), and the 
equivalent of the torus is the hypertorus ($T^3$), which can be 
obtained by identifying opposite faces of a cube. As for the
two-dimensional case, $R^3$ and $T^3$ are locally identical but
globally different.

There exist many other multiply connected 3-D spaces of constant
curvature. These can each be represented by a fundamental 
polyhedron (like the rectangle for the case of $T^2$) embedded 
in the simply connected space of the same curvature
(i.e. in $R^3$, $S^3$ or $H^3$), of which the faces of the polyhedron
are identified in pairs in some way. The simply connected 
space is then the covering space, and can be thought of in 
format (iii) as above, as an apparent space which is tiled 
by copies of the fundamental polyhedron, just as the mosaic
floor of a temple may be (in certain cases) tiled by a repeated
pattern of a single tile.

If the physical Universe corresponds to a multiply connected space 
which is small enough, i.e. which is finite and for which photons
have the time to cross the Universe several times, then the 
(apparent) observable Universe would be a part of the covering
space and would contain several copies of the fundamental polyhedron.
In other words, in apparent space, there could be multiple 
apparent copies of the single physical Universe.

The possibility of seeing several times across the Universe provides
the basic principle of nearly all the methods capable of constraining 
or detecting the topology of the Universe: a single object 
(or a 3-D region of black-body plasma) should be seen in different 
sky directions and at different distances (hence different 
emission epochs). These multiple images, such as 
the three images which, according to the observationally inspired
hypothesis of \citet{RE97}, could be three images of a single
cluster of galaxies (Coma) seen at three different redshifts,
are called {\em topological images}. 

For a thorough introduction to the subject (but prior to the recent surge
in observational projects), see \citet{LaLu95}. For more recent
developments, see \citet{Lum98}
and \citet{LR99}, and workshop proceedings in \citet{Stark98} 
and \citet{BR99}.

\label{s-models}
\section{Friedmann--Lema\^{\i}tre--Robertson--Walker universes 
and their extensions}

So, in order to begin to know the `shape' of the Universe, both
the curvature {\em and} the topology need to be known.

However, virtually all of the observational estimates of cosmological
parameters have been estimates of local cosmological parameters. 
The curvature parameters $\Omega_0$ (present value of the matter 
density parameter expressed in units of the density which would 
imply zero curvature if the cosmological constant is zero) 
and $\lambda_0$ (present value of the dimensionless
cosmological constant) and $H_0$ (the Hubble constant, which
sets a time scale) are each defined locally at a point 
in space.

Estimates of the values of these parameters are now honing in rapidly,
and a convergence from multiple observational methods for the three
parameters is likely to 
signal a new phase in observational cosmology. However,
as explained above, this will leave unanswered
the basic question: how big
is the Universe? Good estimates of the  curvature parameters and of
$H_0$ will help search for cosmic topology, and will constrain the
families of spaces (3-manifolds) 
possible, but will be insufficient to answer the question.

Observations such as the cosmic microwave background (CMB), 
the abundance of light elements and numerous observational statistics
of collapsed objects as a function of redshift lend support 
to the standard FLRW or hot Big Bang model as a good approximation to the
real Universe. According to this model, the age of the Universe
is finite. 

This condemns us to live in an observable universe which is finite,
in which we are situated right at the centre, from the point of view
of the universal covering space. The observable Universe can be defined
as the interior of a sphere (in the covering space) 
of which the radius is the distance
travelled by a photon that takes nearly the age of the Universe
to arrive in our telescopes. The value of this
radius, the horizon
distance, is $c/H_0$ to within an order of magnitude, depending on 
which distance definition one uses and on the curvature parameters.
This explains the common misconception according to which the 
value of $H_0$ sets the size of the Universe. The scale $c/H_0$ is
not the size of the Universe, it is just the order of magnitude 
size of the {\em observable,} non-Copernican Universe.

In comoving coordinates (in which galaxies are, on average, stationary,
where the expansion of the Universe is represented by a multiplicative
factor $a$), and using the `proper distance' 
(Weinberg, 1972, eq.14.2.21), the horizon radius is in the range
$6000${\hMpc}$ < R_H \ltapprox $12000{\hMpc} for a range of 
curvature parameters including those which at least some cosmologists
think are consistent with observation\footnote{$h\equiv 
H_0/(100$km s$^{-1}$ Mpc$^{-1}$).}.

Note that the observable Universe is very non-Copernican: we are
at the centre of a spherical Universe. Of course, the underlying model
implies that the complete covering space is (probably) much larger: finite
for positive curvature, infinite for non-positive curvature, 
and in neither case does the covering space have a centre.

Note also that the 2-sphere $S^2$ does not have a centre which is part
of $S^2$.  The centre of a 2-sphere embedded in $R^3$ exists in $R^3,$
but is not part of the 2-sphere. $S^2$ can be very easily defined as a
mathematical object independently of $R^3$. The embedding in $R^3$ is
certainly a useful mathematical 
tool, and an aid to intuition, but is not at all
necessary.  So, if $S^2$ corresponds to a physical object, this does
not imply that $R^3$ has physical meaning, nor that the `$R^3$-centre'
of $S^2$ has any physical meaning. 
The exactly corresponding arguments apply to $S^3$ relative to $R^4$.

If Robertson and Walker's implicit hypothesis that the topology
of the Universe is trivial were correct (the hypothesis according
to which, for example, the 3-torus $T^3$ is a priori excluded),
then, since the observations seem to indicate that the Universe is 
either negatively curved (hyperbolic) or flat, not only would the
covering space be infinite, but the Universe itself would be!
This would imply that the fraction of the Universe which is 
observable would be zero, since the observable
Universe is finite. It would also imply (for a constant average
density, the standard assumption) that the mass of the Universe
is infinite.

This may or may not be correct. Atoms have finite masses, as 
do photons, trees, people, planets and galaxies. If the Universe
is a physical object, then extrapolation from better known physical objects
would suggest that it should also have a finite mass.

Both theoretical and observational
methods can be used to examine the hypothesis of trivial topology.

Many theoretical cosmologists and 
physicists work on extensions to the standard 
model, to epochs preceding that during which the 
cosmic microwave background black-body radiation was emitted
(e.g. see the early universe, topological defect and superstring
cosmology papers in \citealt{BhKar00}). 
Inflation (an accelerated expansion of the Universe at an
early epoch, e.g. when the age of the Universe was $\sim 10^{-33}s$)
and other theoretical ideas regarding the `early' Universe don't
invalidate the standard Big Bang model as a good approximation for
post-recombination observations (i.e. probably all observations so far),
even if some now include `no Big Bang' boundary conditions at
the quantum epoch $t \sim 10^{-43}s$. On the contrary, they 
extrapolate from the standard model.

Among these various scenarios, some treat the Universe as having 
infinite volume, some as finite, and many do not state either way.

If we consider one of the early Universe models in which the volume
is infinite or, else, say, the Universe is globally a hypersphere 
with radius $10^3$ times that of the horizon, and 
if we assume that the topology of the Universe is trivial, 
then a more or less serious question of credibility arises: 
is the extrapolation from the observable Universe to the entire 
Universe $10^3$ times or infinitely many times bigger justified? 
Is an extrapolation from an `infinitesimal' (i.e. zero) fraction 
to the whole justified?

Whether these questions lie in the domain of physics or of
the philosophy of science will not
be dealt with further here, except to remark that for the sake of
precision, it would be best to make it clear in literature for
the non-specialist 
when one is studying the `observable Universe' or the `local Universe',
and not leave the term `Universe' without an appropriate qualifying
adjective.

It is clear that if the topology is assumed to be trivial, then
the measured values of local parameters such as $\Omega_0$ and 
$H_0$ would be `local' in more than one sense of the word: local
as a physical quantity, and local since the values are averaged
over an `infinitesimal' fraction or, say,  
a ten billionth of the total volume of the Universe.

How does the assumption of trivial topology relate to the quasi steady
state cosmology model, which is a model of many `mini' Big Bangs
averaging out to a constant density (in space and time) universe?
Trivial topology seems to be an implicit (though probably not
necessary) assumption of the model.
The zero curvature version provides a universe model which
is globally infinite in both space and time if the topology is
trivial, without any preferred epochs, satisfying the `perfect
cosmological principle'.  If observations significantly
showed that the topology of the Universe were non-trivial, i.e. if
photons were shown to have `wrapped' many times and in different
directions around the Universe in less than its present age, then this
simplest version of the quasi steady state model would have significant
problems: the Universe would be finite in at least one (spatial) direction.

If a quasi steady state model (of any curvature) were multiply
connected, then a characteristic length scale would exist.  If this
scale were observable at the present, despite the overall exponential
expansion of the model since an infinite past (an overall hyperbolic
sine or hyperbolic cosine contraction and expansion in the curved
models), then this would imply that we happen to live at a special epoch 
in the infinite history of the Universe, which would contradict
the original motivations for these models.

One possible solution might be for topological evolution to occur at
the minima of each short time scale expansion cycle, so that at least
one closed geodesic is visible during each cycle. If the whole
fundamental polyhedron is found to be observable, then a model in
which the universe snaps off into several independent fundamental
polyhedra (universes) at the minimum of each cycle might be sufficient
to match the observations. However, topological change would
presumably require quantum effects, i.e. would require the Universe to
be dense enough to go through a Planck epoch (where quantum mechanics
and general relativity both need to be applied) at each cycle
minimum. Since one of the motivations for the quasi steady state model
was the avoidance of the conventional explanation 
of the cosmic microwave background
as photons coming from a horizon scale high density state, the introduction 
into the model of a global, much higher density state 
would again be problematic.

%

\label{s-obsvns}
\section{Dropping the simple-connectedness hypothesis}

The hypothesis that the Universe is simply connected is {\ldots} 
just a hypothesis. 

If this hypothesis is dropped, then the whole Universe may well 
be {\em smaller} than the `observable Universe'! The latter would
then form a part of the universal covering space, and would 
constitute the `apparent Universe' containing many copies of the
entire physical Universe. Multiple connectedness does not necessarily 
imply that multiple copies would be visible (one or all dimensions
might be bigger than the horizon diameter), but certainly implies
this as a physical possibility.

As mentioned above, awareness that measurement of topology would be 
required in order to characterise the geometry of space has been
around for at least a century \citep{Schw00}, and has been discussed
by several of the symbols of modern cosmology 
\citep{deSitt17,Fried24,Lemait58}. Although measuring curvature, 
essentially via estimates of the density parameter, $\Omega_0$,
and the cosmological constant, $\lambda_0$,
has sustained much more attention and observational analyses than
measurement of topology, some discussion of the latter both
theoretically and in relation
to the status of continually growing observational catalogues 
of extragalactic objects was made in the 1970's and 1980's, 
in particular by Ellis, Sokoloff and Schvartsman, Zel'dovich, 
Fang and Sato, Gott and Fagundes [see \cite{LaLu95} for a detailed
reference list, also, e.g. \citet*{NarSesh85,BJS85}].

Since the release of data from the COBE satellite, 
several papers were quickly published to make statements 
about spatial topology with respect to the COBE data
\citep{Stev93,Sok93,Star93,Fang93,JFang94}. The publication
of a major review paper \citep{LaLu95} further prompted
interest in the subject, so that 
there are now 
several dozen researchers in Europe, North America, Brazil, China, 
Japan and
India actively working on various observational 
methods for trying to measure the topology of the Universe.

\fhowcirtwo

\fcircles

See \citeauthor{LR99} (1999, Section 5) for a detailed discussion 
of the recently developed observational methods, apart from new
work which is cited below. For earlier work, which showed by various
methods that the size of the physical Universe should be at least
a few 100{\hMpc}, see \citet{LaLu95}.

Most of the methods depend either directly or indirectly 
on multiple topological imaging of either collapsed
astrophysical objects or of photon-emitting regions of plasma.

Other methods are the statistical incompatibility
between observable topological defects and observable cosmic topology 
\citep{UzPet97,Uzan98,Uzan98b}, and the 
suggestion of \citet{Rouk00b} which
postulates a physical and geometrical link between the 
$L \approx 130${\hMpc} feature in large scale 
structure and global topology.

The direct methods are those for which 
photons are expected to 
travel across the Universe in different directions from
a single object or plasma region and arrive at a single observer. 
They may leave the object or plasma region at different cosmological
times. 

The indirect methods suppose that regions which are nearby
to one another have correlated physical properties, so that although
an object or plasma region is not strictly speaking multiply imaged,
close by regions are approximately multiply imaged. This approach
is subject to the validity of the assumptions regarding correlations
over `close' distances.

\subsection{Three-dimensional methods (collapsed objects)}

Researchers in France and Brazil
(\citealt{Fag87,Fag96,Fag98}; 
Lehoucq, Luminet \& Lachi\`eze--Rey 1996;
Roukema 1996; Roukema \& Blanloeil 1998;
 Roukema \& Bajtlik 1999; Roukema \& Luminet 1999;
\citealt*{Gomero99a,LLU98,ULL99,FagG97,FagG99a,FagG99b,Gomero99b,Gomero99c}) 
work principally on direct three-dimensional methods, i.e. study various
statistical techniques which analyse the spatial positions of
all known astrophysical objects at large distances inside of 
the observational sphere. Just as for traditional observational
estimates of the local cosmological parameters $\Omega_0, H_0$ 
and $\lambda_0,$ the idealised methods have to be adapted in practice 
to cope with the fact that we observe objects in the past and with 
astronomical selection effects.
Different classes of objects have different constraints on their
evolution with cosmological time, are seen to 
different distances and are observed in a combination of wide
shallow surveys and narrow deep surveys. The result is that the
search for cosmic topology --- or claimed proofs of simple connectedness
below a given length scale --- are just as difficult as the attempts
to measure the curvature parameters and the Hubble constant, efforts
which have taken more than half a century in order to start coming
close to convergent results.

One way to find a very weak signal in a population such as quasars
which are likely to evolve strongly over cosmological time scales 
can be seen schematically in Fig.~\ref{f-topo4}. This is the
search for rare local isometries \citep{Rouk96}. Although the
properties of individual quasars may have changed completely between
the high redshift and low redshift images, the relative 
three-dimensional spatial positions of a configuration of quasars 
should remain approximately constant in comoving coordinates.
Even if such isometries are rare, use of a large enough catalogue
may be enough to detect enough isometries to generate testable
3-manifold candidates, whose predictions of multiple topological
imaging can be tested by other means.

In a population composed of good standard candles and negligible
selection effects, the `cosmic crystallography' method of \citet{LLL96}
(or its variations) 
could be applied. 
By plotting a histogram of pair separations (in the covering space) 
of the objects (i.e. an 
unnormalised two-point correlation function), sharp `spikes' should
occur at distances corresponding to the sizes of the vectors representing
the isometries between the copies of the fundamental polyhedron 
{(\em generators)}. 
The original version of cosmic crystallography 
is valid only for flat spaces. However, \cite{ULL99} 
extrapolated the search for local isometries [discussed and 
calculated for quintuplets in \citet{Rouk96}] to
the case of isometries of pairs, and defined a single statistic 
based on the number of `isometric' pairs. This 
`collecting-correlated-pair' statistic should 
have a high value in the presence of multiple topological imaging 
in a population composed of good standard candles, 
whether space is hyperbolic, flat or spherical, but only for the
correct values of the curvature parameters $\Omega_0$ 
and $\lambda_0$.

Most of the three-dimensional methods avoid having to make 
an a priori hypothesis regarding the precise 3-manifold (space),
its size and orientation. Since there are infinitely many 
3-manifolds possible, this is a considerable advantage for 
hypothesis testing. However, if carried out to sufficient precision
and generality 
to guarantee a detection, in the case that an observational data
set is homogeneous and deep enough and the topology of the Universe
really is non-trivial, the methods generally require a lot of computing
power, generally in CPU rather than in disk space.

Various ideas to improve the speed of the calculations are suggested
by some of these authors. Alternatively, if some observations can
be used to suggest candidate 3-manifolds (spaces), then the methods 
can be used on different, independent observational data sets to 
attempt to refute the suggested candidates relatively rapidly.

\subsection{Two-dimensional methods (cosmic microwave background)}

Both direct and indirect approaches 
have been suggested for using 
CMB measurements, those of COBE and future measurements by MAP 
and Planck Surveyor.

\subsubsection{The direct approach: the identified circles principle}

The direct method results from an elegant geometrical result
discovered by a North American group 
(Cornish, Spergel \& Starkman 1996, 1998). This is known as
the identified circles principle. 

The microwave backgrounds seen by two observers in the
{\em universal covering space} separated by a
comoving distance $R$ lying in the range $0 < R < 2R_H$ consist of 
two spheres in the covering space, which intersect in a circle
in the covering space. If the two observers are multiple topological
images of a single observer, then what is a single circle in 
the covering space as seen by two observers is equivalent to two
circles seen in different directions by a single copy of the
observer (see Fig.~\ref{f-howcir2}).

This implies that for a single observer, 
and for locally isotropic radition on the surface of
last scattering, the temperature fluctuations seen around a circle
centred at one position in a CMB sky map should be identical 
to those seen around another circle (in a certain direction), apart
from measurement uncertainty. The positions and radii of the 
circles are not random, they are determined by the shape, 
size and orientation of the fundamental polyhedron, or equivalently,
by the set of generators.

\citet{Corn98b} intend to use the MAP satellite data to apply this
principle in a generic search for the topology of the Universe.
It has been shown that the principle can
be applied to four-year 
data from the COBE satellite despite its poor resolution 
and poor signal-to-noise ratio,
either to refute a given topology candidate motivated by 
three-dimensional data \citep{Rouk00a} or to show that 
a flat `2-torus' ($T^2 \ttimes R$) model a tenth of the horizon
diameter can easily be found which is consistent with the COBE 
data (see 
Fig.~\ref{f-circles2} here, or \citealt{Rouk00c} for details).

For the identified circles principle to be applied in a general way,
i.e. to search for the correct 3-manifold rather than to test a
specific hypothesis, the computing power required
would again be very high,
as for the three-dimensional methods. Improvement in the speed of 
calculation will be required for the application to MAP and Planck
Surveyor data.

Another direct method, which has been tested to some degree via
simulations, and which should in principle be derivable from the
identified circles principle, is that of searching for patterns 
of `spots' in the CMB \citep{LevSGSB98}.

\subsubsection{The indirect approach: use of
perturbation statistics assumptions}

Many researchers 
\citep{Stev93,Sok93,Star93,Fang93,JFang94,deOliv95,dOSS96,LevSS98}
have tried to use the indirect approach, i.e. via introducing 
assumptions on the density perturbation spectrum (or equivalently 
the correlation function of density perturbations) and making 
statements regarding ensembles of possible universes, as opposed
to direct observational refutation. Most of these authors made 
simulations of the perturbations in order to obtain statements
of statistical significance. 
Various subsets of flat 3-manifolds were tested 
and it was suggested that flat 3-manifolds
up to 40\% of the horizon diameter were inconsistent with the
COBE data. As mentioned above, direct tests applying the identified
circles principle show that a more conservative constraint would
have to be around 10\% of the horizon diameter.

A Canadian based group \citepf{BPS98,BPS00a,BPS00b} has tested
individual hyperbolic candidates applying the perturbation statistics
approach to COBE data. Since Fourier power spectra are, 
strictly speaking, incorrect in hyperbolic space, and since eigenmodes
are difficult to calculate in compact hyperbolic spaces, these 
authors used correlation functions instead.
Although these authors use some simulations in their figures, 
the perturbation statistics approach is applied by them without relying on
simulations. This bypasses potential errors and numerical limitations
which could be introduced by the transition from perturbation statistics
to simulations to final statistical statements (though it does not
avoid the original assumptions).

Inoue (1999), Aurich (1999)  and \citet{CS99} calculated eigenmodes
in compact hyperbolic spaces in order to apply the perturbation 
simulational approach. They showed that the $C_l$ (spherical harmonic)
statistic for COBE four-year data was consistent with their models,
taking into account the fact 
that for low density universe models, i.e. for $\Omega_0 < 1$, 
the gravitational redshifts between the observer and the surface
of last scattering, known as the integrated Sachs-Wolfe effect (or
the Rees-Sciama effect) makes refutation of 3-manifold candidates
using CMB data more difficult than if the Universe were flat and the
cosmological constant were zero. 

Applications of the perturbation statistics approach (with or
without simulations) to testing
multiply connected models have the
property that they depend on the assumptions regarding density
fluctuation statistics. The latter are observationally supported by
most (but not all) analyses of COBE data on large scales if the
Universe is assumed to be simply connected, which may not be a valid
assumption if the Universe is multiply connected.
For more discussion on this question, see Section 1.2 of \citet{Rouk00a}.

\section{Summary and Prospects}

Is it possible to show by observations that the global Universe 
is observable? And that observational cosmology is something more 
than just extragalactic astrophysics?

For a recent summary of observational results, see table of \citet{LR99}.
The analyses using different observational 
data sets and different methods
have so far answered these questions with `No' on scales which 
are much smaller than the horizon.
All the observations
point to the Universe being simply connected up to a scale 
of $\sim 1000${\hMpc}, i.e. about a tenth of the horizon diameter. 

At the $\sim 1000${\hMpc} scale and larger, 
definitive answers have not yet been obtained.
On the contrary, some candidate 3-manifolds consistent with several
observational data sets have been suggested
by \citeauthor{RE97} \citep{RE97,RBa99,Rouk00a}, 
\citet{Rouk00c} and \citet*{BPS98,BPS00a,BPS00b}.

Specific testing of these large 3-manifolds
might show that one of these makes correct 
predictions of three-dimensional positions (celestial coordinates
and redshift) of multiple topological images of previously known objects.

In parallel, developments of all the techniques described above are
being actively pursued by the different groups cited above, using
computer simulations and different classes of observations.  

Strong hopes are put in the MAP satellite which should be launched in
the next year or two to map the CMB, 
but the better resolution and the ability to
measure polarisation information in the CMB 
by the Planck Surveyor might be
needed to extract the signal from the noise.

Using three-dimensional methods,
numerous surveys such as the 2dF and SDSS galaxy and quasar surveys
which are presently underway 
might be sufficient to obtain significant results.

Theory may be slow in catching up. Quantum cosmology studies regarding
the evolution of topology during the quantum epoch are nowhere
near making predictions for the present-day topology of space,
though some interesting work has begun (e.g. \citealt{DowS98}).

Not only would a significant measurement of cosmic topology show
that the Universe is spatially finite in at least one direction,
but it would add topological `lensing' to the tool presently
finding a great variety of useful applications: gravitational lensing.
The two are similar in that the geometry of the Universe generates
multiple images in both cases. They differ in that the former uses
the whole Universe as a `lens', does not magnify the image and
generates images at (in general) widely differing redshifts and
angles, all of which contrast with the latter.

\noindent{\bf Acknowledgments}

This review is partially inspired from that of \citet{Rouk99},
and the author thanks Jean-Philippe Uzan, Varun Sahni and
Tarun Souradeep for providing numerous useful comments.


\def\apj{ApJ.}                        
\def\apjs{ApJ. Supp.}                 
\def\aj{A.J.}                           
\def\aanda{A\&A}                      
\def\cqg{Class. Quant. Grav.}         
\def\mnras{Month. Not. R. Astron. Soc.}

\newcommand\joref[5]{#1, #5, {#2, }{#3, } #4}
\newcommand\epref[3]{#1, #3, {#2}}


\end{document}